# The electronic origin of shear-induced direct to indirect gap transition and anisotropy diminution in phosphorene

Baisheng Sa[1,*], Yan-Ling Li[2,*], Zhimei Sun[3], Jingshan Qi[2], Cuilian Wen[1] and Bo Wu[1]

[1]*College of Materials Science and Engineering, Fuzhou University, 350108, Fuzhou, People's Republic of China*
[2]*School of Physics and Electronic Engineering, Jiangsu Normal University, 221116, Xuzhou, People's Republic of China;*
[3]*School of Materials Science and Engineering, and Center for Integrated Computational Materials Engineering, International Research Institute for Multidisciplinary Science, Beihang University, 100191, Beijing, People's Republic of China*

**Abstract:** Artificial monolayer black phosphorus, the so-called phosphorene has attracted global interest with its distinguished anisotropic optoelectronic and electronic properties. Here, we unraveled the shear-induced direct to indirect gap transition and anisotropy diminution in phosphorene based on first-principles calculations. Lattice dynamic analysis demonstrated that phosphorene can sustain up to 10% applied shear strain. The band gap of phosphorene experiences a direct to indirect transition when 5% shear strain is applied. The electronic origin of direct to indirect gap transition from 1.54 eV at ambient condition to 1.22 eV at 10% shear strains for phosphorene was explored and the anisotropy diminution in phosphorene is discussed by calculating the maximum sound velocities, effective mass and decomposed charge density, which signals the undesired shear-induced direct to indirect gap transition in the applications of phosphorene for electronics and optoelectronics. On the other hand, the shear-induced electronic anisotropy properties suggest that phosphorene can be applied as the switcher in the nano electronic applications.

*To whom all correspondence should be addressed: bssa@fzu.edu.cn (B.S. Sa) or

ylli@jsnu.edu.cn (Y.L. Li).

## 1. Introduction

The discovery and investigation of graphene with the two dimension (2D) hexagonal honeycomb carbon promotes the emergence of 2D materials [1, 2]. Apart from graphene, many novel 2D materials, for instance, elementary silicene and germanene, functional graphene, silicene and germanene [3, 4], binary single layer boron-nitride [3] as well as monolayer transition metal dichalcogenides [5] have attached global interests. The fascinating and unexpected electronic, optical, mechanical, magnetism and thermal properties of these low dimensional materials provide unlimited possibilities of artificial materials and potential applications. For example, functionalized germanenes were predicted to be topological insulators with finite 2D band gaps and 1D gapless edge states [6]. Monolayer boron-nitride [3], $MoS_2$ [7] as well their related 2D van der Waals heterostructures [8-10] were designed as the photocatalysts in the applications of visible light induced water splitting hydrogen production.

Very recently, monolayer black phosphorus or phosphorene has been fabricated [11-13] and attached increasing interest due to its peculiar anisotropic optoelectronic and electronic properties [14-20]. Potential applications as nano-electronic transistors [12], solar cells [18] and photocatalyst [21] have been proposed for this novel mono-elementary 2D material, which shows high carrier mobility [22-24], negative Poisson's ratio [25] and enormous mechanical flexibility [26]. Many studies show that the electronic structure of phosphorene is very sensitive to the external conditions. For instance, the geometric, stabilities and diffusions of the monovacancy and divacancy in phosphorene have been calculated [27], the mechanism of semiconductor to metal transition [28] and direct to indirect band gap transition [29] in phosphorene under applied inter-planar strains have been theoretically explored. Recently, Fei *et al.* found that the preferred conducting direction in phosphorene can be rotated by appropriate intra-planar biaxial or uniaxial strains [30]. All these previously studies are focused on phosphorene under the applied single-axial or biaxial strains. However, the study of phosphorene under shear strain, another very important and commonly used external condition for 2D materials [31, 32], is still lacking. Considering strain engineering is one of the most important material processing technologies in the semiconductor industry [33], a comprehensive study of shear strained phosphorene is of great interest and importance as well.

In this work, we systematically studied the crystal and electronic structures of phosphorene under intra-planar shear strains based on the density functional theory calculations. The lattice dynamical stability of phosphorene under certain shear strain and the sound speed as well as structure deformation were discussed. We further unraveled the electronic origin of the direct to indirect gap transition in phosphorene under shear strain. Moreover, the effective mass and decomposed charge density were further studied to uncover the anisotropy diminution in phosphorene.

## 2. Computational details

The density functional theory calculations of phosphorene were based on the Vienna *ab initio* simulation package [34] (VASP) in conjunction with projector augmented wave (PAW) pseudopotentials within the generalized gradient approximations [35] (GGA) of Perdew-Burke-Ernzerhof [36] (PBE). The valence electron configuration for P was $3s^23p^3$. The geometry convergence was achieved with the cut-off energy of 500 eV. K-points of 8×10×1 were automatically generated with the Γ symmetry. In order to prevent the interaction between periodic images, we introduced a vacuum of 20 Å thickness in the *z*-axis direction. The relaxation convergence for ions and electrons were $1\times10^{-6}$ eV. The PHONOPY [37] code was applied to get the phonon frequencies through the supercell approach with a 3×6×1 supercell and 2×2×1 K-points. Since the van der Waals interactions in the phosphorene are less significance than the bulk black phosphorous, the van der Waals corrected optB88 function [38] was applied for comparison only. The Heyd-Scuseria-Ernzerhof (HSE06) hybrid functional [39] with 25 % short range exchange energy from Hartree-Fock was introduced to evaluate the band gap of phosphorene under shear strains. The iso-surfaces of the decomposed charge density were visualized using the VESTA [40] tool.

**3. Results and discussions**

The sketch of the crystal structure of phosphorene is presented in Figure 1a, phosphorene has an orthogonal 2D crystal structure. The relaxed lattice parameters $a$ = 4.626 Å and $b$ = 3.298 Å agree well with previously studies [29, 30]. The distortion of the lattice is defined by multiplying the Bravais lattice vectors matrix with the applied shear distortion matrix $\aleph_s = \begin{pmatrix} 1 & \varepsilon_s \\ \varepsilon_s & 1 \end{pmatrix}$, where $\varepsilon_s$ is the applied shear strain. Figure 1b shows the schematic diagram of the phosphorene lattice under applied shear strain.

Firstly, we explored the stability of phosphorene under applied shear strains. Figure 2 illustrated the phonon dispersion curves of phosphorene under various shear strains. As can be seen in Figure 1a, our calculated phonon dispersion curves for phosphorene without shear strain agrees well with the previously studies [41]. Interestingly, even 1% shear strain splits off all the double degenerated phonon dispersion modes along the X to M and M to Y directions. Meanwhile, the double degenerated modes at the X and Y points remain degenerated. However, the splitting of the original degenerated modes does not break the stability of the phosphorene lattice. It can be found that the phosphorene lattice remains stable under up to 10% applied shear strain without any imaginary mode (Figure 1b-c). As the applied shear strain increased into the range from 11% to 15%, there exhibits *fcc* ThH$_2$-type [42, 43] imaginary acoustic modes along the Γ to X direction and the Γ to Y direction, showed in Figure 2d-e. Although such imaginary modes can be enhanced to positive by the temperature effects or temperature induced electron phonon interactions [44]. Such negative dispersion modes indicate the lattice instability of phosphorene lattice under certain shear strains. By further increasing the applied shear strain to 16% (Figure 2e), one double degenerated acoustic phonon mode become negative around the Y to M direction indicates strong lattice instability, which may lead to the crash of the phosphorene lattice. Since the application of phosphorene must be based on good stability, we will analyze the physical and chemical properties of phosphorene under no more than 10% shear strain in the following part.

For phosphorene, there are four atoms in one unit cell with three acoustic phonon modes and nine optic phonon modes in the phonon dispersion curves. The optic phonon frequencies at the Γ point are classified according to the irreducible representations. The even parity vibrations $B_{1g}$, $B_{3g}^1$, $A_g^1$, $B_{3g}^2$, $B_{2g}$ and $A_g^2$ with the inversion symmetry at the Γ point are Raman active modes; and the odd-parity vibrations $B_{1u}$, $A_u$ and $B_{2u}$ showing anti-symmetry at the Γ point are infrared (IR) modes. More details about the atomic motions of vibrational modes can be found in Fei *et al.*'s [45] letter. The calculated phonon frequencies at the Γ point for phosphorene under shear strain are plotted in Figure 3. Interestingly, the optic modes behave different under applied shear strain. The frequencies of rigid $B_{1g}$ and $B_{3g}^1$ modes only slightly decreased when applying the shear strain. But the frequencies of $B_{1u}$, $A_g^1$, $A_u$ and $B_{2g}$ modes gradually decreased. Meanwhile, a significant red shift can be observed in the soft $B_{3g}^2$ mode. All these red shifts come from the reduced interatomic interactions corresponding to the dispersion modes in phosphorene under shear strain [45]. At the same time, obvious blue shifts can be found in the $A_g^2$ and $B_{2u}$ modes, which originate from the increasing interatomic interactions corresponding to the dispersion modes [45].

We have further studied the sound velocities by fitting the slopes of the acoustic dispersion curves around the Γ point. At ambient conditions, our calculated maximum longitudinal velocities of sound along the Γ to X direction $v_{max}^{\Gamma-X}$ = 3.96 km/s and along the Γ to Y direction $v_{max}^{\Gamma-Y}$ = 7.99 km/s, which agree well with previously results by Zhu *et al*. [46] Herein, $v_{max}^{\Gamma-Y}$ is about twice as much as $v_{max}^{\Gamma-X}$, indicating strong lattice dynamic anisotropy in phosphorene. Figure 4 illustrated the sound velocities as a function of the applied shear strain. Obviously, $v_{max}^{\Gamma-X}$ will be increased with the increase of the applied shear strain. Meanwhile, the $v_{max}^{\Gamma-Y}$ will be slightly decreased. The maximum velocity of sound along the Γ to X direction is $v_{max}^{\Gamma-X}$ = 6.81 km/s and the Γ to Y direction is $v_{max}^{\Gamma-Y}$ = 7.56 km/s under 10% shear strains. As a result, the applied shear strain reduces the strong anisotropy of lattice dynamic properties in phosphorene.

The details of the crystal structure of phosphorene can be found in our previously work [21]. To explore the crystal structure change of phosphorene under shear strains, we plotted the P-P bond length and bond angle as a function of the applied shear strain in Figure 5. As can be seen in the figure, the intra-planar P-P bond length $R_1$ will be slightly decreased with the increase of the shear strain, the variation of $R_1$ is 0.023 Å from 0 to 10% shear strain. Meanwhile, 10% shear strain introduces only 0.002 Å increasing for P-P bond length $R_2$. Hence, the P-P bonds show very tiny response to the shear strain. At the same time, as the shear strain increases, the intra-planar bond angle $\theta_1$ keeps its value around 96°, which do not response to the shear strain as well. While the inter-planar angle $\theta_2$ drops sharply from 104.18° to 95.78° under 0 to 10% shear strains. Hence, the applied shear strain changes the phosphorene lattice by mainly decreasing the inter-planar angle $\theta_2$.

Phosphorene has been predicted to be a nearly direct band gap semiconductor near the Γ point, and can be briefly considered as a direct gap semiconductor [28, 47]. It is well-known that PBE usually underestimates the band gap for semiconductors, while the hybrid functional with the mixing of the Hartree-Fock and DFT exchange energy is a practical method to exactly evaluate the band gap. Herein, we plotted the band gap for phosphorene under shear strain both using the PBE and HSE06 methods [39] in Figure 6. We have obtained 1.54 eV band gap using HSE06 and 0.91 eV band gap using PBE for phosphorene. We considered the nearly direct band gap nature of phosphorene at ambient condition as the direct gap at Γ. We found that phosphorene keeps the direct band gap behavior though its band gap is slightly decreased under no more than 4% applied shear strain, which is similar to the effect of certain compression strain [29]. By further increasing the applied shear strain, phosphorene will be transformed to an indirect band gap semiconductor, and the value of the indirect band gap decreases sharply with the increase of shear strain. As can be seen in Figure 6, HSE06 hybrid functional enlarges the band gap without any further significant change of the electronic structure, where PBE and HSE06 method represent very similar features and tendencies for the band gap under shear strain. Hence we will only present the further analysis of the electronic band structure of phosphorene under shear strain based on PBE in the following.

Phosphorene shows very strong electronic anisotropy features along the Γ to X and Γ to Y directions [22]. Figure 5a shows the strain free band structure of phosphorene. Similar to the phonon dispersion curve, the electronic bands along the X to M and M to Y directions are double degenerated. To further analysis the electronic nature of the direct to indirect band gap transition of phosphorene under shear strain, we plotted the electronic band structure of phosphorene under different shear strains. By comparing the band structures under different shear strains, we found that the shear strain has remarkable effects on the band structure along the X to M and M to Y directions, but not the Γ to X and Γ to Y directions. Since the bands along the Γ to X and Γ to Y directions are very sensitive to the axial strains [29, 30], different behavior of the phosphorene band structure under applied shear strain is expected. As can be seen in Figure 7b, 1% shear strain splits off all the double degenerated bands along the X to M and M to Y directions. At the same time, the double degenerated states are pinned at the X and Y points. As a result, corresponding shear strain does not change the electronic features of phosphorene along the Γ to X and Γ to Y directions. The splitting along the X to M and M to Y directions increases with the applied shear strain increasing. By increasing the applied shear strain to 5%, the conduction band minimum (CBM) shifts from the $p_z$ electron dominated state at the Γ point to the *s* electron dominated state locates in the middle of the M to Y direction, which comes from the splitting of the *s-p* hybridized bands [21]. Meanwhile, the valence band maximum (VBM) remains at the Γ point. As a result, phosphorene will transfer from a direct gap semiconductor to an indirect gap semiconductor. Further increasing of the applied shear strain to 10% will not change the indirect band gap feature. It is worth noting that the *s* dominated CBM in the middle of the M to Y direction under 5% to 10% shear strain presents more homogeneous electronic characterizations, which may reduce the anisotropy in phosphorene.

To evaluate the electronic anisotropy in phosphorene under shear strain, we calculated the effective masses of the electron (hole) by fitting the states around CBM (VBM) [48]. Our calculated effective masses of the electron along $x$ direction $m^*_{e,x}$ = 0.148 $m_e$ and $y$ direction $m^*_{e,y}$ = 1.237 $m_e$. Our calculated effective masses of the hole along $x$ direction $m^*_{h,x}$ = 0.138 $m_e$ and $y$ direction $m^*_{h,y}$ = 5.917 $m_e$. We can define the effective mass anisotropy according to $A_e = \dfrac{m^*_{e,y}}{m^*_{e,x}}$, and $A_h = \dfrac{m^*_{h,y}}{m^*_{h,x}}$. For an ideal isotropic electronic system, the values of $A_e$ and $A_h$ equal to 1, while any value smaller or greater than 1 is a measure of the degree of electronic structure anisotropy possessed by effective masses along different direction. Herein, $A_e$ = 8.36 and $A_h$ = 42.88 for phosphorene without applied any shear strain. These results indicate strong electronic anisotropy of phosphorene and agree well with previously studies [22, 29, 30]. Figure 8 illustrated the effective mass of electron and hole in phosphorene under different applied shear strains. As can be seen, under no more than 4% shear strain, the electron effective mass along $x$ direction keeps its value while the electron effective mass along $y$ direction grows with the applied shear strain increasing. Interestingly, the electron effective mass jumps suddenly along $x$ direction under 5% applied shear strain, which drops sharply along $y$ direction. Such phenomenon originates from the direct-indirect band gap transition during the CBM shifting we have discussed above. As a results, the effective masses anisotropy $A_e$ = 0.41 under 5% applied shear strain, which will further increase to $A_e$ = 0.64 under 10% applied shear strain. On the other hand, there is no VBM shifting during the direct-indirect band gap transition, hence the hole effective mass varies gradually. As the shear strain increases, the hole effective mass along *x* direction keeps its value while the hole effective mass along *y* direction gradually decreases. As a results, the effective masses anisotropy is decreased to $A_h$ = 20.41 under 10% applied shear strain. Thus applied shear strain remarkably reduces the electron effective mass anisotropy and partially reduce the hole effective mass anisotropy in phosphorene. Since the carrier mobility is inversely dependent on the effective masses, the carrier mobility anisotropy in phosphorene can be adjusted by applied shear strain as well.

To further explore the electronic anisotropy, we studied the decomposed charge density of the CBM of phosphorene. Figure 9a plots the iso-surface of decomposed charge density of the Γ point CBM of shear strain free phosphorene in real space, which is very similar to the previously studies [30]. Continuous electron charge clouds dominate the space above and below the phosphorene layer along *y* direction. On the other hand, discontinuous electronic charge clouds are divided into individual parts along *x* direction inside the phosphorene layer. Hence the CBM of phosphorene shows a strong anisotropy feature. However, according to the CBM shifting during the direct-indirect band gap transition, the decomposed charge density of the CBM of phosphorene under 5% applied shear strain presents a more homogeneous character, which is shown in Figure 9b. Continuous electron charge along both *x* and *y* direction fabricates a 2D charge net inside the phosphorene layer. This result agrees well with our previously analysis about the electron effective mass by fitting the states around VBM.

One of the most important reasons why phosphorene have been attracted global interest in the potential applications for electronics and optoelectronics is that phosphorene is a direct band gap semiconductor. Its band gap, band alignment and work function can be tuned by changing sample thickness [22, 49]. Considering phosphorene can theoretically hold up to 30% critical strain [26], and its physical and chemical properties can be tuned by strain engineering [28, 30]. Strained phosphorene should be widely used in the electronic and optoelectronic devices like strained-silicon MOSFETs [50] in the near future. According to our analysis, there exists a direct to indirect band gap transition in phosphorene under shear strain as small as 5%. Hence, any form of shear strain, for example, global shear strain, process-induced shear strain or post-processing shear strain, should be avoided in the applications of strained phosphorene. Herein, the anisotropy diminution of phosphorene can be the sign of the appearance of the undesirable shear strain. From another point of view, the electronic anisotropy properties of phosphorene, for instance, the carrier mobility anisotropy, can be adjusted by the shear-induced direct to indirect gap transition, which indicate that phosphorene can be applied in the potential application as the shear strain protected nano electronic switcher.

## 4. Conclusion

In summary, we have systematically studied phosphorene under applied shear strain based on the density functional theory calculations. We found that phosphorene is dynamically stable under no more than 10% applied shear strain. Further analysis of the sound velocities indicates that the applied shear strain reduces the strong anisotropy of lattice dynamic properties in phosphorene. We studied the direct to indirect band gap transition and unrevealed the related conduction band minimum shifting in phosphorene under shear strain. Furthermore, the anisotropy diminution of the electronic structure of phosphorene was explored by investigating the effective mass as well as decomposed charge density. Our result reveals that strong shear strain should be avoided when designing engineering phosphorene-based electronic or optoelectronic devices. However, phosphorene shows potential application as the nano electronic switcher with its shear strain sensitive anisotropy properties.

**Acknowledgments**

This work was supported by the National Natural Science Foundation of China (Grant Nos. 11347007, 61274005, 51301039 and 51171046), the National Natural Science Foundation for Distinguished Young Scientists of China (Grant No. 51225205), Qing Lan Project, and the Priority Academic Program Development of Jiangsu Higher Education Institutions (PAPD).

**Figure captions**

Fig. 1. (a) The structure sketch of phosphorene. (b) The schematic diagram of the applied shear strain according to Eq. (2).

Fig. 2. The phonon dispersion curves of phosphorene under (a) 0 %, (b) 1 %, (c) 10 %, (d) 11 %, (e) 15 %, (f) 16 % shear strain.

Fig. 3. The frequencies of optical modes at the $\Gamma$ point of phosphorene under applied shear strain.

Fig. 4. The sound speed of phosphorene as a function of the applied shear strain.

Fig. 5. (a) The P-P bond length and (b) bond angle as a function of the applied shear strain.

Fig. 6. The band gap of phosphorene as a function of the applied shear strain.

Fig. 7. The band structures of phosphorene under (a) 0 %, (b) 1 %, (c) 5 %, (d) 10 % shear strain.

Fig. 8. The electron and hole effective mass of phosphorene as a function of the applied shear strain.

Fig. 9. The decomposed charge density plots of the conduction band minimum of phosphorene under (a) 0 % and (b) 5 % shear strain.

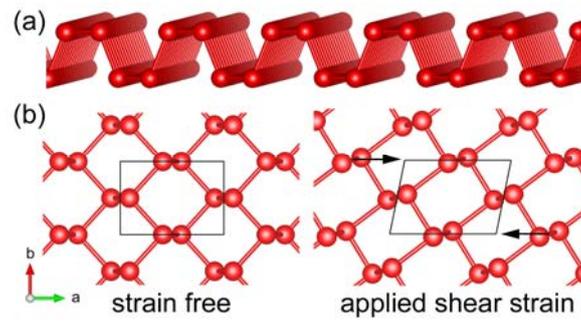

Fig. 1. (a) The structure sketch of phosphorene. (b) The schematic diagram of the applied shear strain according to Eq. (2).

Fig. 2

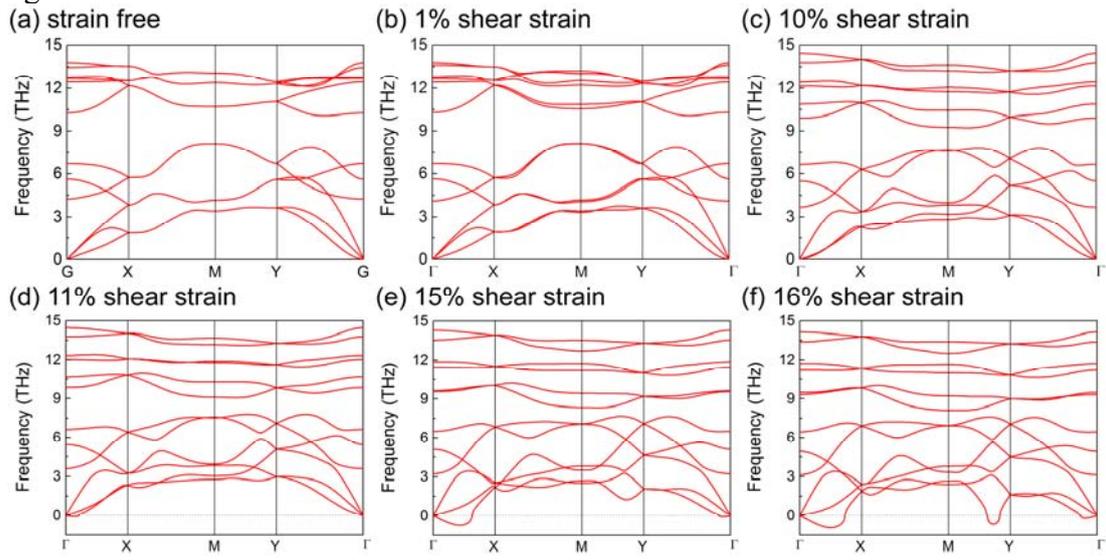

Fig. 2. The phonon dispersion curves of phosphorene under (a) 0 %, (b) 1 %, (c) 10 %, (d) 11 %, (e) 15 %, (f) 16 % shear strain.

Fig. 3.

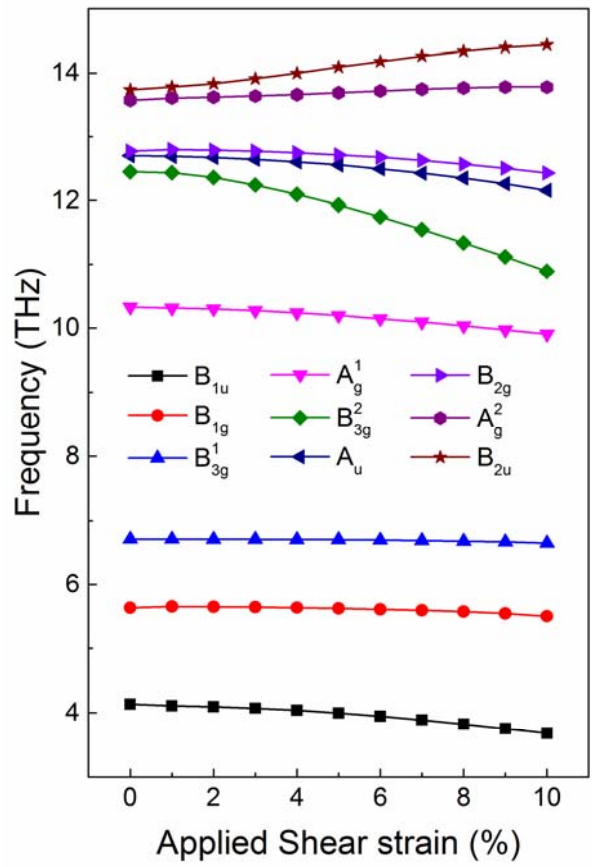

Fig. 3. The frequencies of optical modes at the Γ point of phosphorene under applied shear strain.

Fig. 4.

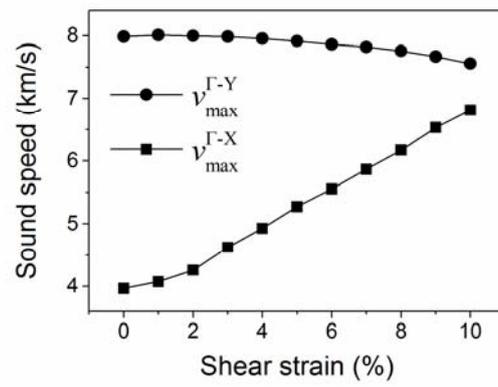

Fig. 4. The sound speed of phosphorene as a function of the applied shear strain.

Fig. 5.

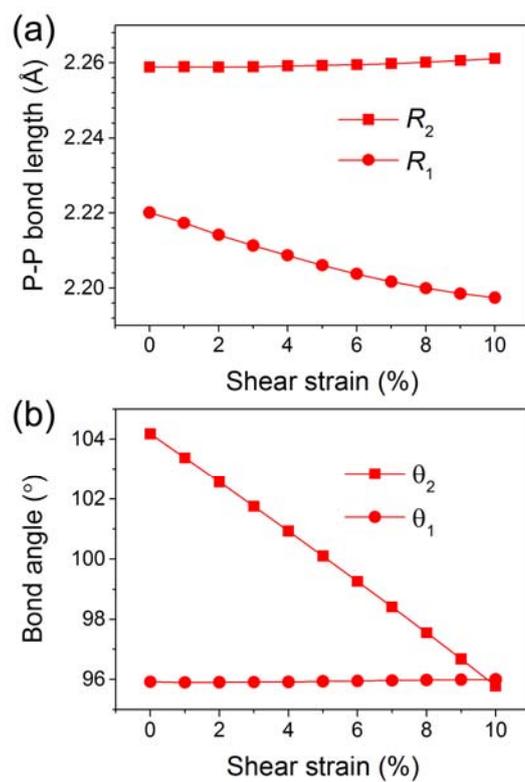

Fig. 5. (a) The P-P bond length and (b) bond angle as a function of the applied shear strain.

Fig. 6.

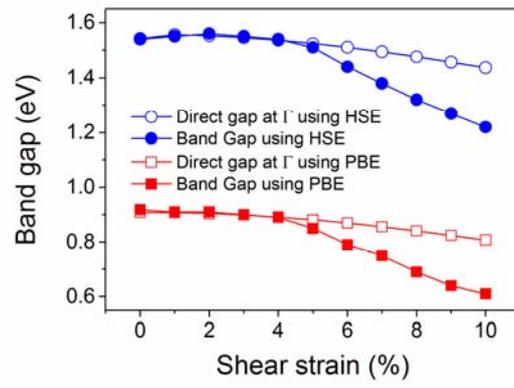

Fig. 6. The band gap of phosphorene as a function of the applied shear strain.

Fig. 7.

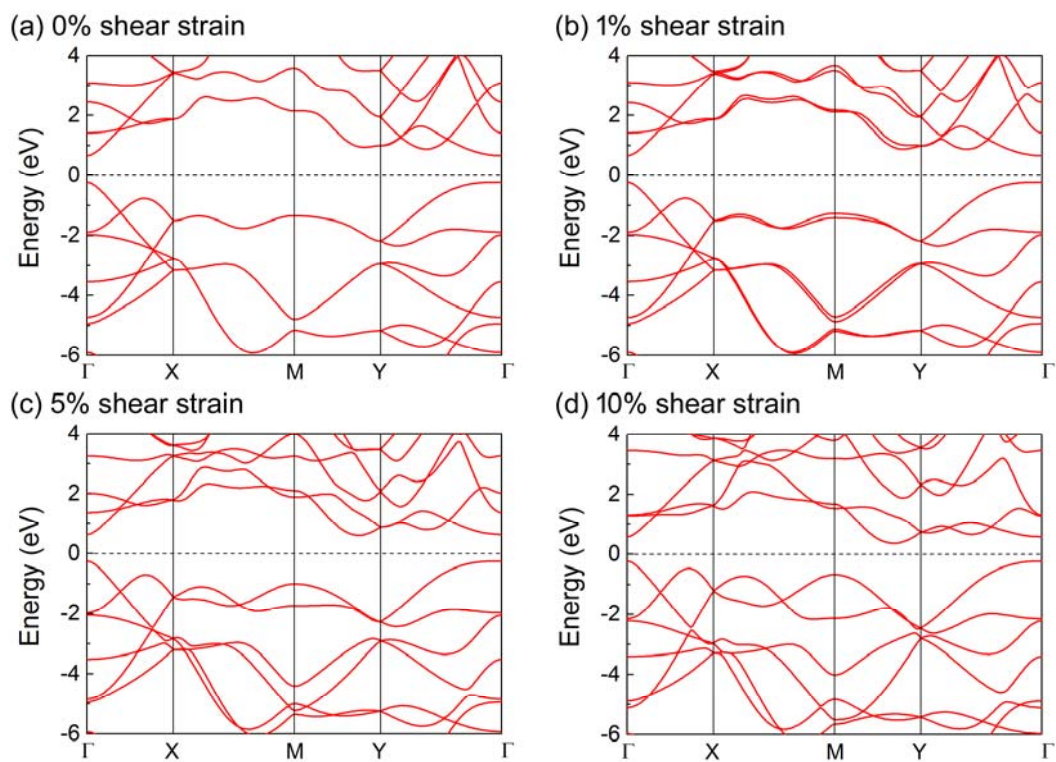

Fig. 7. The band structures of phosphorene under (a) 0 %, (b) 1 %, (c) 5 %, (d) 10 % shear strain.

Fig. 8.

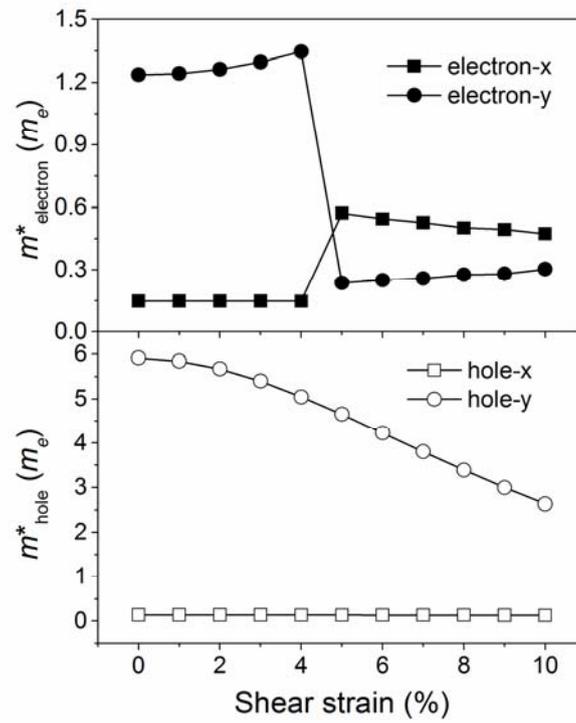

Fig. 8. The electron and hole effective mass of phosphorene as a function of the applied shear strain.

Fig. 9.

(a) strain free

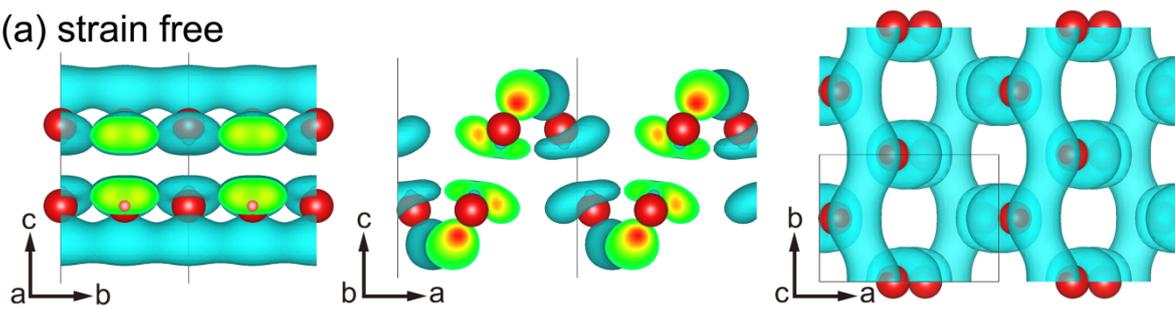

(b) 5% shear strain

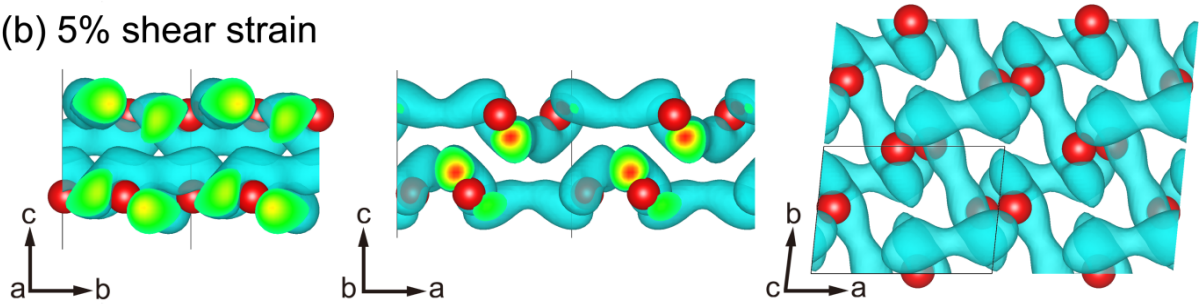

Fig. 9. The decomposed charge density plots of the conduction band minimum of phosphorene under (a) 0 % and (b) 5 % shear strain.